\documentclass[5p,10pt,times]{elsarticle}
\usepackage{epsfig}
\usepackage{amssymb}

\begin{document}

\begin{frontmatter}



\title{Cross section versus time delay and trapping probability}

\author{G. A. Luna-Acosta\footnote{Tel: +52 (222) 2295610, Fax: +52
(222) 2295611 (G. A. Luna-Acosta)}}
\ead{glunaacosta@gmail.com}
\address{Instituto de F\'isica, Benem\'erita Universidad Aut\'onoma de
Puebla, Apartado Postal J-48, Puebla 72570, Mexico}

\author{A. A. Fern\'andez-Mar\'in}
\address{Instituto de Ciencias F\'isicas, Universidad Nacional Aut\'onoma de M\'exico, 
A.P. 48-3, 62251 Cuernavaca, Morelos, Mexico}

\author{J. A. M\'endez-Berm\'udez}
\address{Instituto de F\'isica, Benem\'erita Universidad Aut\'onoma de
Puebla, Apartado Postal J-48, Puebla 72570, Mexico}

\author{Charles Poli}
\address{Department of Physics, Lancaster University, Lancaster, LA1 4YB, United Kingdom}

\begin{abstract}
We study the behavior of the $s$-wave partial cross section $\sigma(k)$, the 
Wigner-Smith time delay $\tau(k)$, and the trapping probability $P(k)$ as function of the 
wave number $k$. The $s$-wave central square well is used for concreteness, simplicity, 
and to elucidate the controversy whether it shows true resonances. It is shown that, 
except for very sharp structures, the resonance part of the cross section, the trapping 
probability, and the time delay, reach their local maxima at different values of $k$. We 
show numerically that $\tau(k)>0$ at its local maxima, occuring just before the resonant 
part of the cross section reaches its local maxima. These results are discussed in the light 
of the standard definition of resonance.
\end{abstract}


\end{frontmatter}

\section{Introduction }

By far, the most widely used scattering function is the cross section $\sigma(k)$. Its
analysis provides essential information about all kinds of scattering phenomena in physics
and it is specially important in the study of resonances. For sharp resonances, the 
{\it resonance part} of the cross section is generally assumed to be described by the 
famous Breit-Wigner resonance formula. Its importance cannot be overestimated since
this formula is given in terms of the parameters that characterize the resonance; namely,
the width and position of the (cross section) resonance (see e.g.~\cite{Workman2013} 
and references therein). This is perhaps the reason why,
quite often, the term resonance is taken to mean a resonance in the cross section. In fact,
it is sometimes explicitely stated that the resonance energy is defined as that which
corresponds to the value of $\pi/2$ of the resonant part of the phase shift, see 
e.g.~\cite{Descouvemont2010}. Furthermore, these parameters can be related to some
fundamental quantities. For example, in nuclear physics, the width (position) corresponds
to the decay width (mass) of a meta-stable particle \cite{Ceci2013}. 
Just as knowledge of individual resonances is important, so is the understanding of their 
statistical properties, such as the distribution of their widths or spacings, specially in the 
field of quantum chaos \cite{BDietz2010}.  

Since the cross section is defined in terms of the phase shift $\theta(k)$ and
this is composed of a background part and a resonance part, to unveil the sought after
resonance parameters from the data a complicated fitting procedure must be applied
\cite{Workman2013}.
Thus, study of other scattering functions can yield important and complementary
information about the system. 

The Wigner-Smith delay time $\tau(k)$ \cite{Carvalho2002} is such a quantity. Actually,  
knowledge of $\tau(k)$ may be considered necessary in order to comply with the most 
generally accepted definition of resonance \cite{Taylorbook,Newtonbook,ABohmbook}. 
Namely, a rapid increase in the phase shift through $\pi/2$ (modulo $\pi$). How fast? 
Fast enough so that there is time delay since a time delay implies the existence
of a meta-stable state and viceversa \cite{ABohmbook}.

In the literature resonances are sometimes defined as the poles of the scattering matrix 
$S$ \cite{Taylorbook,ABohmbook,PDG2014,LL65}
(for a more mathematical definition see \cite{Simon00}).
We shall refer to these as resonance {\it poles} to distinguish from 
the {\it scattering} resonances discussed above. Certainly, the two definitions are connected and the scattering resonances may be viewed as the manifestation of the resonances poles (ocurring near the real axis). However, while the definition of resonance poles is unambiguous, the definition of scattering resonances seems to lack certain consistency that we shall try to point out in the remainder of this paper.

In this paper we shall focus on comparing the behavior of three scattering functions:
The $s$-wave partial cross section $\sigma(k)$, the Wigner-Smith time delay $\tau(k)$, 
and the trapping probability $P(k)$ (to be defined in the next section).
We will show that the ``speed of the phase shift" $l(k)\equiv 2(\partial \phi/\partial k)$,
see Eq.~(\ref{eq.2p7}) below,
can be considered itself a scattering function and it is the link between the other three
functions.

One of the objectives of this work is to show that different closely related scattering 
functions do not in general  peak at the same $k$ values as the resonant part of the cross 
section and hence their study offers  complementary information about the resonance 
properties of the system under study.  Although the differences in the peak positions of 
the various scattering functions may be small there are underlying conceptual differences 
that may lead to a deeper understanding of resonance phenomena. We shall see that the 
centers of the resonances of $l(k)$ and $\tau(k)$ can be identified with the real part of 
the $S$-matrix poles in $k$-space, whereas those of $\sigma_{\phi}(k)$ with the absolute 
value of the pole.
 
To be able to get exact results for these quantities we shall use the $s$-wave central
square well potential. Despite the simplicity of this potential, it has served not only as a
textbook example to display basic features of quantum resonance scattering 
\cite{Weberetal1982, Nussenzvieg59} but also as a model for some nuclear systems, see 
e.g.~\cite{Descouvemont2010,Lane86,Vogt1962,BlattWeisskopf}.
Paradoxically, some authors have pointed out that the square well does not give rise to 
``true" non-zero energy resonances \cite{Taylorbook,Meyer76}. The reason being that
precisely at resonances of the cross section, the time delay is at most equal to zero. 
Others maintain that in spite of this the square well does produce Breit-Wigner resonances, 
thus advocating the ``less restrictive definition of a resonance as an enhancement in the 
cross section due to a pole in the scattering amplitude" \cite{Weberetal1982}. It appears 
then that the study of other scattering functions may elucidate the controversy and 
perhaps induce some polishing in the definition of resonance.

The speed of the phase shift $l(k)$ certainly plays a fundamental role. For the 
central square well, characterized by the ``strength'' $\alpha$ 
(given in Eq.~(\ref{eq.3p1b})), we demonstrate that the local maxima of $l(k)$ occur 
just before the resonant part of the cross section reaches its local maxima. Further, we 
show that $\tau(k)>0$ at all the local maxima of $l(k)$ except for very large values of 
$\alpha$, where $\tau(k)\rightarrow 0$.  

\section{Scattering functions and resonances}

In this brief presentation of the basic scattering quantities, we consider non-relativistic
spinless scattering off finite-range potentials. Specifically, potentials decaying as $1/r$ or slower and short range potentials plus a Coulomb-like potentials are not considered here.
For central finite-range scattering potentials with free-particle asymptotics, the asymptotic
radial wave function $\psi_{\ell}$ for the orbital angular momentum $l$ is 
(see e.g.,~p. 437 in \cite{ABohmbook} or p. 6 in \cite{BurkeAMOBook2011})
\begin{equation}
\psi_{\ell}(k;r) = e^{-i\delta_\ell}+S_\ell(k) e^{i\delta_\ell} , \quad r>a_\ell \ ,
\end{equation}
where $a_\ell$, is the interaction radius for the $\ell$-wave, $\delta_\ell=kr-\ell\pi/2$,
$S_\ell(k)=-e^{2i\theta_\ell(k)}$ is the $\ell$-wave element of the $S$-matrix, and 
$\theta_\ell(k)$ is the $\ell$-wave phase shift. $S_\ell$ can be written in terms of 
$\psi_\ell$ and its space derivative, evaluated at some $r\geq a_\ell$:
\begin{eqnarray}
S_\ell&=& -e^{-2i\delta_\ell}\frac{1+ ik\psi_\ell/\psi'_\ell}{1- ik\psi_\ell/\psi'_\ell} \\
&=&-e^{2i\theta_\ell} \ , 
\end{eqnarray}
where 
\begin{equation}
 \theta_\ell=-(kr-\ell\pi/2) + \phi_\ell 
\end{equation} 
and
\begin{equation}
\phi_\ell=\tan^{-1}[k\psi_\ell(k;r)/\psi'_\ell(k;r)]
\label{phiell}
\end{equation}
is the so-called resonant part of the phase shift. Clearly, the phase shift $\theta_\ell(k)$ 
is independent of the radius $r$ as long as it is in the asymptotic regime ($r \geq a$,
where $a$ is the interaction radius), whereas the resonant part $\phi_\ell(k)$ depends 
on the radius $r \geq a$ where it is calculated. For well defined radius of interaction $a$
the natural choice for the evaluation of $\phi$ is $r=a_\ell$, see also 
Ref.~\cite{Vogt1962}.

\subsection{Cross section}

As shown in most books on quantum mechanics, for central potentials the $s$-wave 
partial cross section is given by
\begin{equation}
\sigma_\ell(k)=\frac{4\pi (2\ell+1)}{k^2}\sin^2(\theta_\ell) \ .
\end{equation}
In this work we shall consider only the case $\ell=0$;  $s$ wave scattering. Dropping the 
subscript $\ell=0$ in all quantities, the $s$ partial wave cross section is then written as
\begin{equation}
\sigma(k)=\frac{4\pi}{k^2}\sin^2(\theta)=\frac{\pi}{k^2}\mid 1+S\mid^2, 
\label{sigmak2}
\end{equation}
where $S=-\exp{(2i\theta)}$, $k$ is the wave number in the asymptotic region, and
$\theta(k)=-kr+\phi$. Clearly, the phase shift $\theta(k)$ is independent of the radius 
$r$ as long as it is in the asymptotic regime ($r \geq a$, where $a$ is the interaction 
radius), whereas the resonant part $\phi$ depends on the radius $r \geq a$ where it is 
calculated. For $s-$waves, if $r=a$, the phase $-ka$ is the so-called hard sphere shift.
As is customary \cite{Vanroose97} and convenient for our purposes, we shall be 
considering the scaled version of the cross section (or scattering amplitude):
\begin{equation}
\label{eq.2p2}
\sigma_\theta(k)=\frac{k^2}{\pi}\sigma(k)=\mid 1+S\mid^2=4\sin^2(\theta)
\end{equation}
and its resonance part
\begin{equation}
\label{eq.2p3}
\sigma_{\phi}(k)=4\sin^2(\phi) \ .
\end{equation}
  
An important reason to separate the phase shift into resonant and non resonant parts is
that the resonance formula of Breit-Wigner refers exactly to the resonant part of the 
phase shift. Since this is not the usual case, there are formulas, like that of Fano's
resonance shape \citep{Fano61} that can be used to fit and extract the so-called 
Breit-Wigner parameters defining the center and the width of the resonance
\cite{BurkeAMOBook2011,Miroshnichenko10}. These parameters, in the relativistic case, 
provide the (Breit-Wigner) mass and life time of the unstable particles. As far as we know 
\cite{Workman2013} this requires fitting procedures with often several fitting parameters. 
The point is that in many  practical applications, the splitting of the phase shift is needed 
to make sense of the data.

\subsection{Time delay and effective traversal distance}

The time delay is a commonly used quantity to characterize resonances, see 
e.g.~\cite{Taylorbook,Newtonbook,BurkeAMOBook2011,Joachainbook}.
In one dimension it is known also as the Wigner-Smith time delay and for a particle of mass 
$\mu$ with incident momentum $\hbar k$, it is defined as \cite{Wigner1955,TSmith60}
\begin{eqnarray}
\label{eq.2p4}
\tau(k) &=& -i\hbar S^{\ast}\frac{\partial S}{\partial E}=2\hbar\frac{\partial \theta}{\partial E}\\
&=&\frac{\mu}{\hbar k}2\frac{\partial \theta}{\partial k} \ .
\label{eq.2p5}
\end{eqnarray}
It is the difference between the time that a particle spends in the internal region in the
presence of a scattering potential minus the time the particle would spend if there were 
no scattering potential \cite{Amrein07}. $\tau(k)$ is directly connected with the
existence of meta-stable states or the temporary capture of the projectile in the
interaction region \cite{ABohmbook}. As mentioned in the Introduction, the established 
definition of resonance requires that $\tau(k)$ be greater than zero. The quantity 
$2(\partial \theta/\partial k$) was called  the ``retardation stretch" by Wigner
\cite{Wigner1955}. He used it to derive his causality condition, discussed in most books 
on scattering theory (see e.g.~page 103 in \cite{Joachainbook} or page 466 in
\cite{ABohmbook}). By splitting the phase shift into resonant and non-resonant parts 
(\ref{eq.2p4}) becomes
\begin{eqnarray}
\label{eq.2p6}
\tau(k)=\frac{\mu}{\hbar k}[l(k) - 2a] \ ,
\end{eqnarray}
where
\begin{equation}
\label{eq.2p7}
l(k)\equiv 2\frac{\partial \phi}{\partial k} \ .
\end{equation}

In contrast with the retardation stretch, the quantity $l(k)$ is the change with $k$ of the  
purely resonant part of the phase $\phi$. 
We shall refer to $l(k)$ as the speed of the resonant part of the phase shift, or, briefly 
speed of the phase shift. In the case of scattering off finite range potentials with a radius
of interaction $a$, the interpretation of $\tau(k)$ as time delay leads us to interpret 
$l(k)$ as the distance a particle would travel under the action of the potential but 
with constant momentum $\hbar k$. That is, an {\it effective traversal distance}.
Note that even though $\tau(k)$ and $l(k)$ are defined in terms of the asymptotic
quantities $\theta$ and $\phi$, respectively, they give information about the time spent
and effective traversal distance within the interaction region.

\subsection{Trapping probability}

For a localized potential with radius of interaction $a$ the quantity  
\begin{eqnarray}
\label{eq.2p8}
P(k) \equiv \frac{1}{a}\int_{int}|\psi(k;r)|^2dr \ ,
\end{eqnarray}
gives the relative probability of trapping or dwelling in the interaction region. Here,
$\psi(k;r)$ is the scattering wavefunction.  
The denominator $a$ in the definition of $P(k)$ implies normalization with respect to a scattering state with uniform density $|\psi(k;r)|^2=1$ within the
interaction region. Although the normalization is arbitrary, what is important is the behavior of $P$ as a function of $k$.  It can be shown that $P(k)$ can be represented alternatively as
\begin{eqnarray}
\label{eq.2p9}
P(k) = \sum_{\lambda}^{\infty}|A^B_{\lambda}(k)|^2 \ .
\end{eqnarray}
Here $A^B_{\lambda}, \lambda=1,2,3..$ are the coefficients in the expansion of 
$\psi(k;r)$ within the internal region ($0\le r \le a$) in terms of a complete set of 
(reaction) states $\Phi_{B,\lambda}$ defined in the reaction matrix theory of Wigner 
\cite{EPE47} (see also 
e.g.~\cite{Descouvemont2010,BurkeAMOBook2011,LaneThomas58, PMbook}). In the 
reaction matrix theory,  the configuration space is divided in two regions; the internal or 
reaction region ($0\le r\le a)$ and the asymptotic region $r\ge a$. $B$ is the boundary 
value parameter specifying the logarithmic derivative obeyed by the set 
$\left\lbrace\Phi_{B,\lambda}, \ \lambda=1,2,...\right\rbrace$.
  
In order to show the relation between $\sigma_{\phi}(k)$, $P(k)$ and $\tau(k)$, it is
convenient to write the tangent of (\ref{phiell}) as 
$\tan(\phi)=kR_0$, where $R_0 \equiv \psi_\ell(k;r)/\psi'_\ell(k;r)$. $R_0$ is the 
so-called reaction function corresponding to the boundary value parameter equal
to zero \cite{LaneThomas58}. It is straighforward to see that
\begin{equation}
l(k)=\frac{2(R_0+kR'_0)}{1+(kR_0)^2} \ , 
\quad R'_0\equiv \frac{\partial R_0}{\partial k} \ .
\label{eq.2p10}
\end{equation}  
It can also be shown, by expressing the expansion coefficients $A^B_{\lambda}$ in 
terms of $R_0$, that
\begin{equation}
aP=2\frac{kR'_0}{1+(kR_0)^2} \ .
\label{eq.2p11}
\end{equation}
Comparing (\ref{eq.2p10}) with (\ref{eq.2p11}) and writing back $R_0$ in terms of 
$\phi$ gives
\begin{equation}
aP=l(k) - \frac{\sin(2\phi)}{k} \ . 
\label{eq.2p12}
\end{equation}
  
This, together with (\ref{eq.2p6}) gives the relation between $P(k)$, $\tau(k)$ and 
$l(k)$. Expressions (\ref{eq.2p10}) to (\ref{eq.2p12}) are true for all central potentials 
as long the potential can be considered negligible for some $r>a$.

\section{Application to the central square well}

\subsection{Expressions}

The potential is  $-|V_0|$ in $0 < r < a$ and zero everywhere else. The wave number $q$
in the internal region is related to the wave number $k$ in the external region by
\begin{equation}
\label{eq.3p1a}
q^2a^2=\frac{2\mu a^2}{\hbar^2}(E+|V_0|)=k^2a^2+\alpha^2  \ ,
\end{equation}
with
\begin{equation}
\label{eq.3p1b}
\alpha^2\equiv \frac{2\mu a^2}{\hbar^2}|V_0| \ .
\end{equation}

The wave function in the internal region is
\begin{equation}
\label{eq.3p2}
\psi(k,r) = A\sin(qr) \ , \qquad 0\leq r \leq a \ ,
\end{equation}
where
\begin{equation}
\label{eq.3p3}
A = -2i\frac{(k/q)e^{-ika}}{\cos(qa)-i(k/q)\sin(qa)} \ .
\end{equation}
The wave function outside the well is
\begin{equation}
\label{eq.3p4}
\psi(k,r) = e^{-ikr}+S e^{ikr} \ .
\end{equation}
 
By means of the continuity of the wave function, c.f.~(\ref{eq.3p2}) and (\ref{eq.3p4}) 
at $r=a$, it follows that
\begin{eqnarray}
S&=& -e^{-2ika}\frac{\cos(qa)+i(k/q)\sin(qa)}{\cos(qa)-i(k/q)\sin(qa)}\\\label{eq.3p5}
&=&-e^{2i\theta} \ , \label{eq.3p6}
\end{eqnarray}
where
\begin{equation}
\theta=-ka + \phi 
\label{eq.3p7}
\end{equation}
and the resonant part of the phase shift $\phi$ is given by
\begin{equation}
\label{eq.3p8}
\tan(\phi)=k\frac{\tan(qa)}{q} \ .
\end{equation}
That is, for the square well, the backround phase shift is $-ka$, the so-called hard sphere
phase shift. It follows from (\ref{eq.3p4}) that the probability $|\psi(k;a)|^2$ of finding
the particle at the boundary between the internal and external regions is equal to 
$\sigma_{\phi}(k)$:
\begin{equation}
|\psi(k;a)|^2=4\sin^2(\phi)=\sigma_{\phi}(k) \ .
\label{eq.3p9}
\end{equation}
This equation shows that it is $\phi$, not $\theta$, that completely determines the
coupling of the internal system with the projectile. $\sigma_{\phi}(k)$ can be expressed in
terms of the amplitude $A$ of the wave function inside the well using continuity of the
wave function at $r=a$ with (\ref{eq.3p2}) and (\ref{eq.3p9}):
\begin{equation}
\label{eq.3p10}
\sigma_{\phi}(k)=|A|^2 \sin^2(qa) \ ,
\end{equation}
where, c.f.~(\ref{eq.3p3}),
\begin{equation}
\label{eq.3p11}
|A|^2=\frac{4 (ka)^2}{(ka)^2 + \alpha^2 \cos^2(qa)} \ .
\end{equation}

Similarly, using (\ref{eq.3p2}) into (\ref{eq.2p8}) and integrating, $P(k)$ can be written
as
\begin{equation}
\label{eq.3p12}
P(k)=\frac{|A|^2}{2}\left(1-\frac{\sin(2qa)}{2qa}\right) \ .
\end{equation}
It can be verified that this expression can be obtained from the more general results:
(\ref{eq.2p11}) or ({\ref{eq.2p12}). 

Now let us consider $\tau(k)$. The bacgkround shift phase is $-ka$. Hence equation
(\ref{eq.2p6}) becomes
\begin{equation}
\tau(k)= \frac{\mu}{\hbar k}\left(l(k)-2a \right) \ .
\label{eq.3p13}
\end{equation}

After a little algebra using, for example, (\ref{eq.2p12}) and (\ref{eq.3p12}) $l(k)$ can
be written as
\begin{eqnarray}
\frac{l(k)}{2a}&=& \frac{|A|^2}{4}\left(1+\frac{2|V_0|}{k^2}\frac{\sin(2qa)}{2qa}\right)
\label{eq.3p14a} \\ 
&=& \frac{P(k)}{2}+\frac{1}{2}\frac{qa\sin2qa}{(ka)^2+\alpha^2\cos^2(qa)} \ .
\label{eq.3p14b}
\end{eqnarray}

\subsection{Results}

The first result is implied by (\ref{eq.3p13}). It tells us how fast the resonant part of the
phase shift must increase in order for meta-stable states to exist. The condition is 
$l(k)>2a$. 

Several authors have obtained approximate expressions for $\tau(k)$ at the local maxima
of $\sigma_{\phi}(k)$, i.e.~at  $\phi=\pi/2$ modulo $\pi/2$, to show that for the square
well the resonant part of the phase shift increases as fast as the background ($-ka$)
decreases. Therefore there is no time delay and consequently, according to the established
definition of resonance, the local maxima of $\sigma_{\phi}(k)$ are not true resonances,
see e.g.~\cite{Taylorbook,Meyer76}. Here, (\ref{eq.3p14a}) together with (\ref{eq.3p11}) show without approximations
that indeed $l(k)=2a$ at $qa=\pi/2$ modulo $\pi$ (equivalenty, $\phi=\pi/2$ modulo 
$\pi$) and hence $\tau(k)$ is exactly zero at the local maxima of $\sigma_{\phi}(k)$. 
However, as we shall show below, $l(k)\ge 2a$ at its own local maxima, which occur very
nearly where the local maxima of $\tau(k)$ occurs. 

We shall first consider in detail four rectangular wells, specified by their width and depth,
listed in the second column of Table~\ref{Table1}. The reasons for choosing these 
particular parameter values will be clear below. 

\begin{table*}[t]
\centering
\large
\begin{tabular}{|c|c|c|c|c|c|c|c|c|c|c|c|}  \hline    \hline
Well & $a,|V_0|$ & $\alpha $ & $Q_B$& $k^*_1$ & $k^{\tau}_1$ & $k^P_1$  & 
$k^{\sigma}_1$ & $\kappa_1$ & $|K_1|$ & $l(k_1^*)/2a$ \\\hline
I & $2.4,10$ & $10.733$ & $3.91$ & $0.8983$ & $0.8934$ & $0.9990$ & 
$0.9950$ & $0.8994$ & $0.9936$ & $1.0486$       
\\ \hline 
II & $12,10$ & $53.665$ & $17.58$ & $0.9915$ & $0.9915$ & $0.9952$& 
$0.9950$ & $0.9913$ & $0.9949$ & $1.0014$  
\\ \hline
III & $12,10/25$ & $10.733$ & $3.91$ & $0.1797$ & $0.1787$ & $0.1990$ &
$0.1990$ & $0.1799$ & $0.1987$ & $1.0486$
\\ \hline 
IV & $8.7326,10$ & $39.0535$ & $12.931$ & $0.4572$ & $ 0.4570$ & 
$0.4716$ & $0.4714$ & $0.4572$ & $0.4714$ & $1.0153$
\\ \hline
V & $8.7766,10$ & $39.2505$ & $12.994$ & $0.0585$ & $0$ & $0.1407$ &
$0.1406$ & $0.0825$  & $0.1407$ & $1.352$ 
\\ \hline
VI & $8.7546,10$ & $39.1520$ & $12.962$ & $0.3274$ & $ 0.3269$ & 
$0.3475$ & $0.3474$ & $0.3279$ & $0.3473$ & $1.0293$ 
\\ \hline
VII & $8.7987,10$ & $39.3489$ & $13.025$ & $1.7948$ & $ 1.7948$ & 
$1.7991$ & $1.7985$ & $1.7943$ & $1.7983$ & $1.0009$ 
\\ \hline
\end{tabular}
\caption{Parameters and scattering properties for seven representative wells. 
The values of $\phi$ at the first local maximum of $l(k)$ are: 1.33, 1.53, 1.33, 
0.68, 1.39, 1.44, and 1.54 for Wells~I to VII, respectively.}
\label{Table1}
\end{table*}

\begin{figure}[t]
\centerline{\includegraphics[width=\columnwidth]{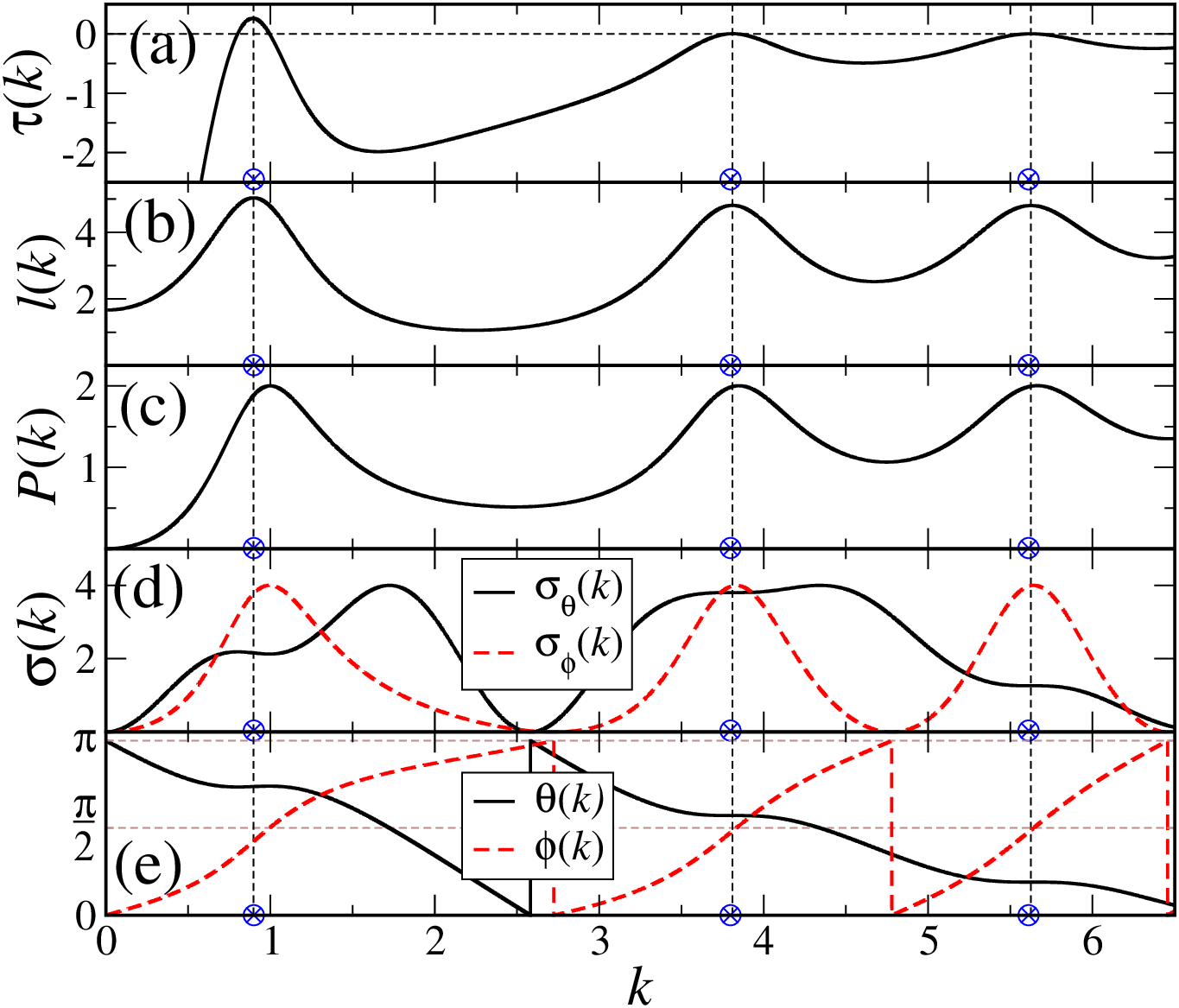}}
\caption{(Color online) Scattering functions for rectangular  Well~I: $a=2.4$, $V_0=-10$, 
and $\alpha=10.733$. (a) $\tau(k)$, (b) $l(k)$, (c) $P(k)$, (d) $\sigma(k)$ 
($\sigma_{\phi}(k)$ and $\sigma_{\theta}(k)$), and (e) the phases $\theta(k)$ and 
$\phi(k)$. The vertical dashed lines indicate the positions of the local maxima of $l(k)$. 
The encircled crosses mark the positions of the real part $\kappa_n$ of the poles of 
the $S$-matrix in complex $k$-space.}
\label{fig1}
\end{figure}

In Fig.~\ref{fig1} we plot the scattering functions $\tau(k)$, $l(k)$, $P(k)$, 
$\sigma_\theta(k)$, and $\sigma_\phi(k)$ for Well~I. In the lowest panel we show 
$\phi$ and $\theta$, modulo $\pi$. In these figures we also mark with an encircled cross 
the positions of the real value of the $S$-matrix pole in $k$-space. These poles were
calculated numerically by finding the complex zeroes of the denominator of $A$ 
(\ref{eq.3p3}). Further, the peaks of $l(k)$ are marked with vertical dashed lines. 
It is evident that not all scattering functions reach their local maxima at the same $k$ 
value. Let us label
the position of the $n^{th}$  peak of $l(k)$, $\tau(k)$, $P(k)$, and $\sigma_\phi(k)$ by
$k^*_n$, $k^\tau_n$, $k^P_n$, and $k_n^s$, respectively.
 
\begin{figure}[t]
\centerline{\includegraphics[width=\columnwidth]{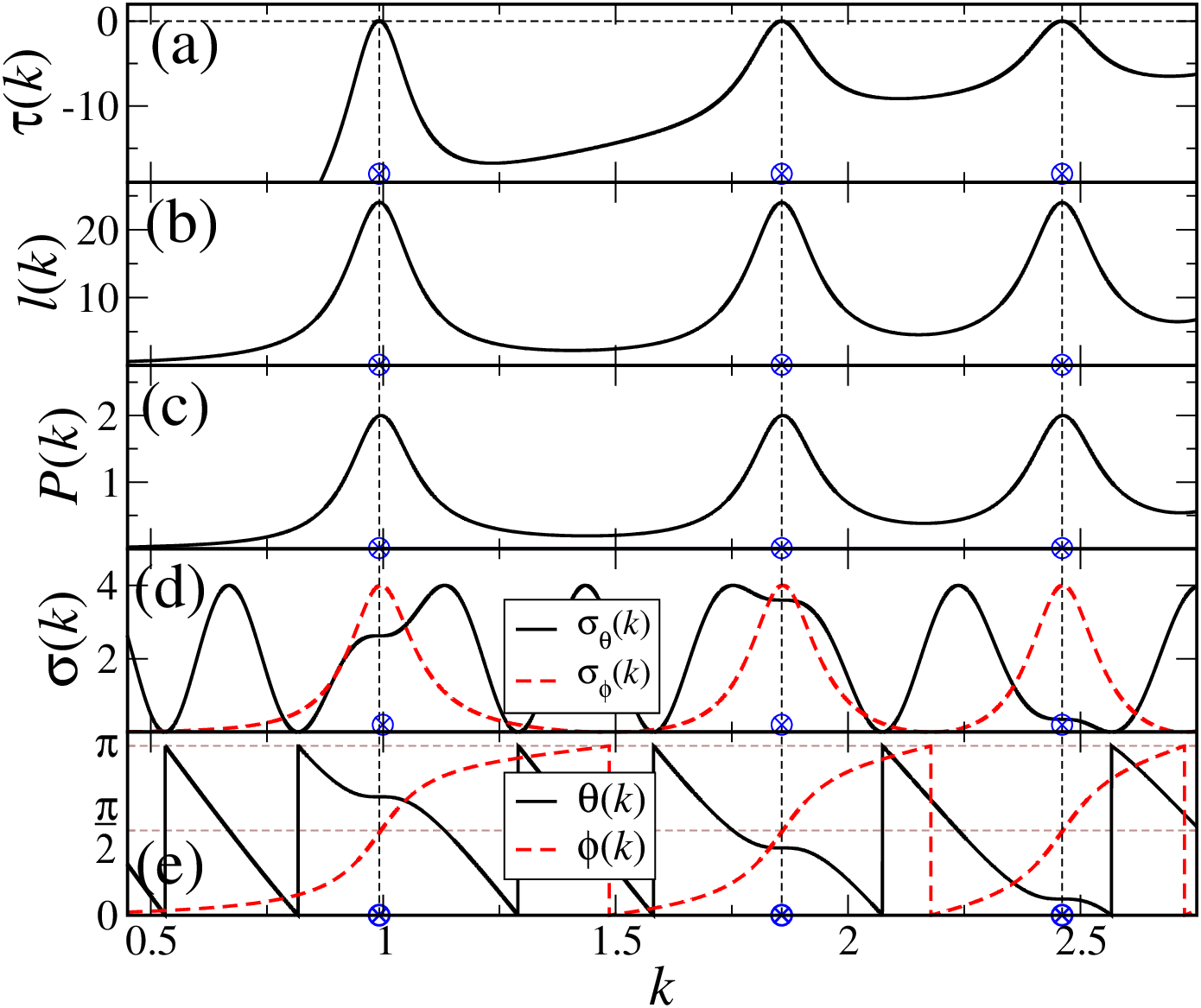}}
\caption{(Color online) Scattering functions for the rectangular Well~II: $a=12$, 
$V_0=-10$, and $\alpha=53.665$. Same description as in Fig.~\ref{fig1}.}
\label{fig2}
\end{figure}
\begin{figure}[t]
\centerline{\includegraphics[width=\columnwidth]{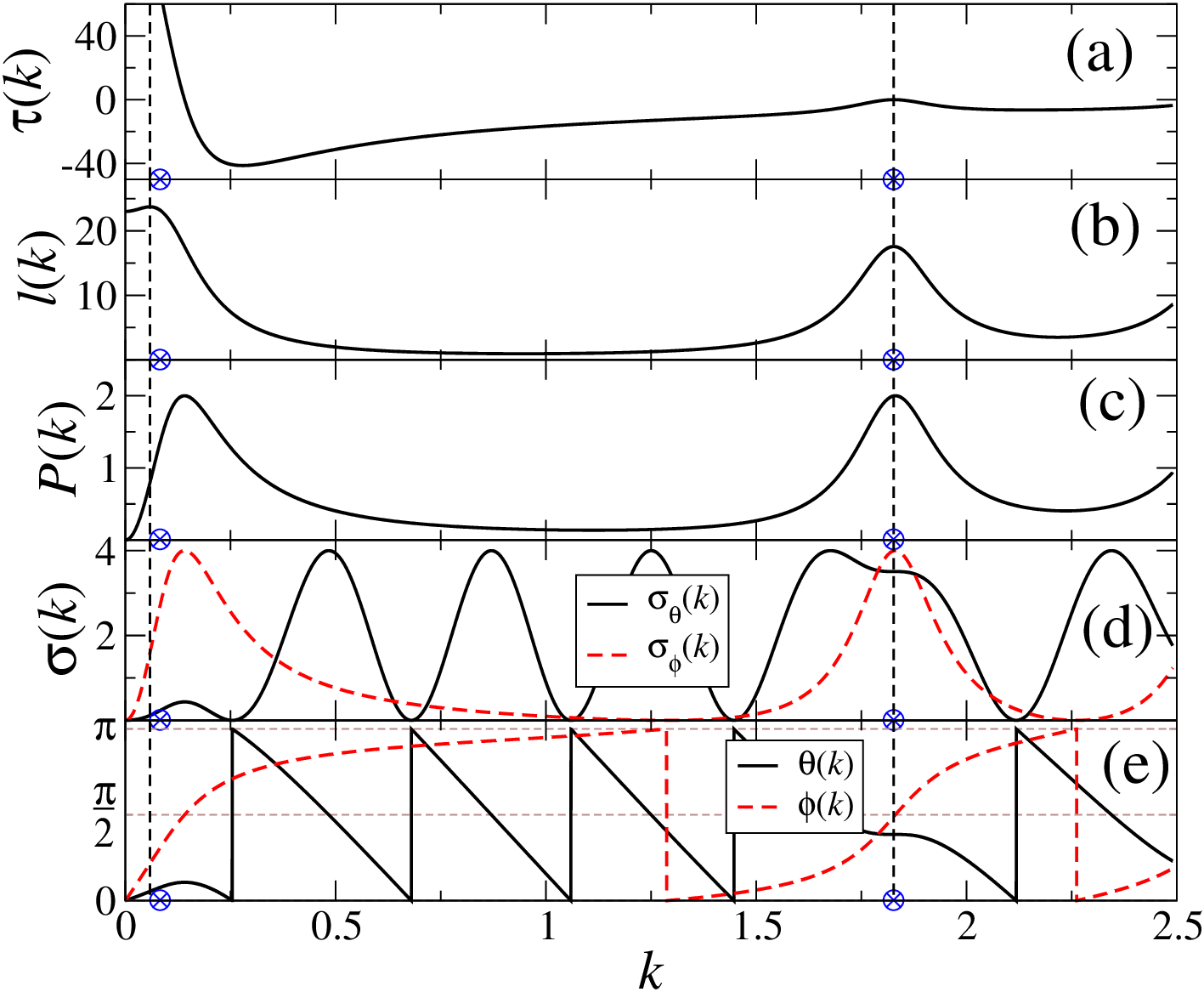}}
\caption{(Color online) Scattering quantities for the rectangular Well~V: $a=8.7766$, 
$V_0=-10$, and $\alpha=39.2505$. Same description as in Fig.~\ref{fig1}.}
\label{fig3}
\end{figure}
\begin{figure}[t]
\centerline{\includegraphics[width=\columnwidth]{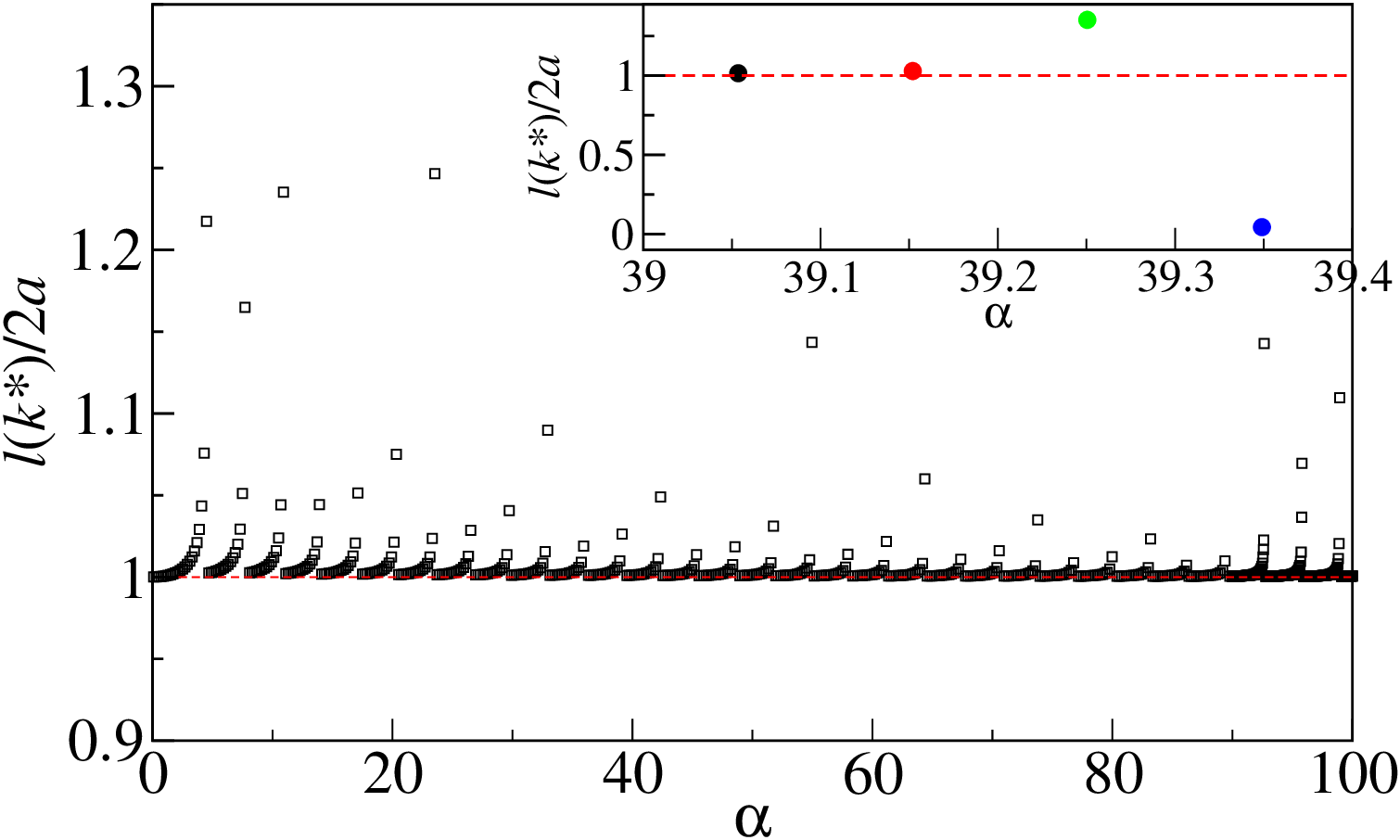}}
\caption{(Color online) First local maxima of $l(k)$ as a function of $\alpha$.
Inset: Details for Wells~IV, V, VII, and VII; see Table~\ref{Table1}.}
\label{fig4}
\end{figure}
\begin{figure}[t]
\centerline{\includegraphics[width=\columnwidth]{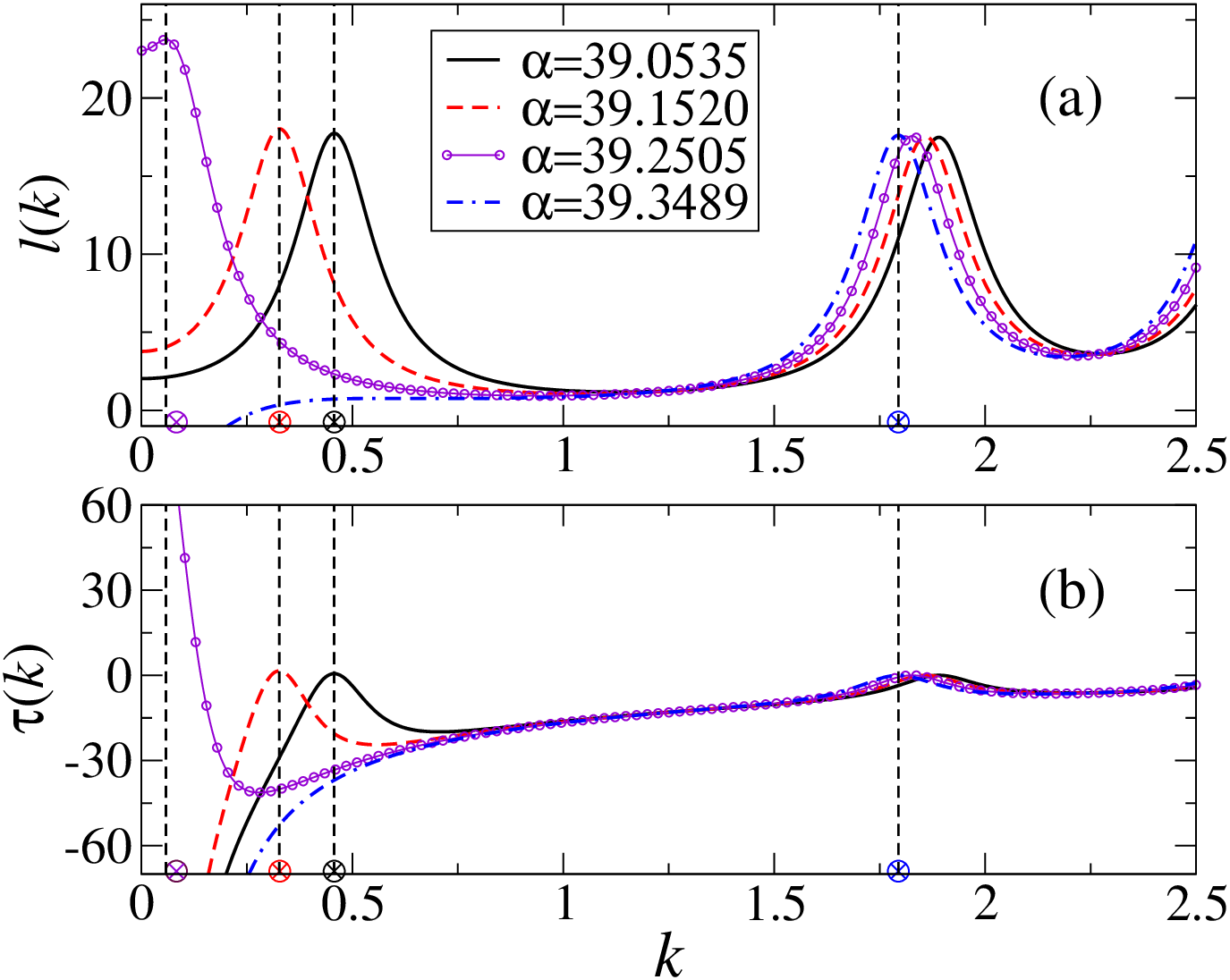}}
\caption{(Color online) (a) Effective traversal length $l(k)$ and (b) time delay $\tau(k)$
for rectangular Wells~IV, V, VI, and VII.}
\label{fig5}
\end{figure}

Let us comment on the following features shown in Fig.~\ref{fig1}:
\begin{itemize}

\item[1.] The peaks of both, $l(k)$ and $\tau(k)$ seem to coincide exactly with the real 
part $\kappa_n$ of the $n^{th}$ resonance $S$-matrix pole in $k$-space. How close 
these are to each other can be seen  by comparing the first row in Table~\ref{Table1}; 
columns 5 and 6 with column 9. The percentual difference between $k_1^*$ and 
$\kappa_1$ is about $0.1$, and  between $k_1^{\tau}$ and $\kappa_1$ is a little larger, 
about $0.6$. 

\item[2.] The centers of the local maxima of $\sigma_\phi(k)$ occur, very close to 
$|K_1|$, the absolute value of the pole in $k$-space. Compare columns 8 and 10, first 
row, Table~\ref{Table1}. The percentual difference is about $0.1$. This may be surprising, 
since it is usually assumed, via the Breit-Wigner formula, that the centers of the 
$\sigma_{\phi}(k)$ peaks correspond to the real part of the $S$-poles in energy space. 
This finding is complementary to that of Klaiman and  Moiseyev \cite{Klaiman2010} who 
showed numerically and analytically that the  transmission resonances for 1D potentials 
(that is 2 channels, instead of one channel as is our case) occur at $|K_n|^2/2$ rather 
than  at its real part $E_R$.

It can be shown analytically, using a one-level model in reaction matrix theory, that indeed  
$k_n^*\approx k_n^{\tau}\approx \kappa_n$ and $k^s_n\approx |K_n|$. The proof of 
these results,valid for all localized central scattering potentials, will be published elsewhere.

\item[3.] The first peak of $P(k)$ seems to occur exactly where $\sigma_{\phi}(k)$ 
peaks (of course, at $\phi=\pi/2$ modulo $\pi$). Table~\ref{Table1} shows that 
actually $k_1^P-k_1^s=0.004 $. The difference $k_n^P-k_n^s>0$ becomes noticeable 
for the second local maxima ($n=2$) and it increases with $n$. This pattern can be 
understood by examining (\ref{eq.3p12}) to (\ref{eq.3p13}) and noticing that all the 
local maxima of both, $\sigma_{\phi}(k)$ and $k^s$, are the same as those of $|A|^2$ 
but different from those of $P(k)$ because of the term $f=1-\sin^2(qa)/(qa)$. For 
$k^P_n$ to be close to $k^s_n$, the term $f$ must be close to one. 
Equation~(\ref{eq.3p12}) shows this occurs for $\alpha\gg 1$ ($\alpha$ of the order of 30 is sufficient) but 
also for intermediate values of $\alpha$ (say, $10$) but for $ka<10$.  In contrast, for 
$\alpha$ of the order of one or less, the term $f$ makes $P(k)$ to be drastically different 
from $|A|^2$. Hence, the behavior as a function of $k$ of the trapping probability and the 
relative intensity (or probability) $|A|^2$ in the interior of the well are very similar 
quantitatively only for very strong potentials ($\alpha\gg 1$) or for intermediate values of 
$\alpha$ but with $ka<\alpha$.

\item[4.] In  panel (a) we see  that both, $\tau(k)$ and $l(k)$ peak apparently at 
$k=\kappa_1$ and $\tau(k)>0$ in a small neighborhood of it. This is remarkable given 
that in panel (d) the full phase $\theta$ looks pretty flat. As we showed above $\tau(k)$ 
is exactly zero at $\phi=\pi/2$ (modulo $\pi$). However, there is a small positive slope, 
just before $\phi=\pi/2$. It is small but sufficiently large to make $l(k)>2a$ (see last 
column of Table~\ref{Table1}) and hence produce a time delay. This positive slope has 
been reported previously \cite{Meyer76} but the authors dismiss by claiming that, in an 
approximation valid for $\alpha$ very large, it cannot create a Breit-Wigner resonance. 
Here we see that it can be considered as a true resonance because the phase shift changes 
fast enough to produce a time delay and, although its maximun does not occur at $\phi=
\pi/2$, it goes through $\pi/2$ during resonance. Moreoever, there is clearly a resonance 
pole associated with the  maximum change in the phase shift and it is responsible too for 
the Breit-Wigner like behavior of $P(k)$ and $\sigma_{\phi}(k)$. We shall determine below 
under what conditions $\tau(k)$ is greater than zero at its local maxima.
 
\item[5.] The plot of the reduced cross section $\sigma_{\theta}(k)$ illustrates how the
background phase shift hides the pure resonance structure evidenced by 
$\sigma_{\phi}(k)$. The effect of the background in the cross section is well known, see 
e.g.~\cite{Taylorbook,ABohmbook,BurkeAMOBook2011}.
For sharp resonances, the effect of the background is studied by writing 
$\sigma_\theta(k)$ in terms of the Fano profile parameter. For not so sharp resonances, 
complicated algorithms must be used and the results are model dependent, as is the case 
in nuclear scattering  phenomena \cite{Workman2013,PDG2014,Anisovich2012}.
\end{itemize}
 
Figure~\ref{fig2} plots the scattering functions of square Well~II. It is 5 times wider than 
Well~I; correspondingly, the ``strength" $\alpha$ is 5 times larger. The resonance 
structure  is sharper than for the Well~I because as $\alpha$ increases the relative 
intensity $|A|^2$ has sharper and deeper oscillations, see~(\ref{eq.3p11}). The effect is 
the same for all scattering functions since they all have the prefactor $|A|^2$ in common. 
The features displayed by this well are qualitatively similar to those of square Well~I except 
that now the peaks of $\tau(k)$ and $l(k)$ move closer to $\phi=\pi/2$. We see that as 
the resonances become sharper the position of the peaks of all these scattering functions 
seem to converge to $k=\kappa_n$, as suggested by Fig.~\ref{fig2}. Another distinctive 
feature is that $\tau(k)$ is now barely larger than zero at its peak. Correspondingly, $l(k)$ 
has increased  by about five times but is barely larger than $2a$, see last column of 
Table~\ref{Table1}.
 
Consider the Well~II listed in Table~~\ref{Table1}. It has different width and depth but 
shares with Well~I the value of $\alpha$. It can be seen from (\ref{eq.3p8}) that $\phi$  
depends on the adimensional variable $ka$ (or $qa$). Then for any two wells with the 
same $\alpha$, the phase $\phi(k)$ corresponding to a well of width $a$ will be the same 
as the phase $\phi(k')$ corresponding to a well of width $a'$ if $ka/a'=k'$. Hence, the 
plots for $P(k)$ and $\sigma_{\phi}(k)$ will be the same but the $k$-axis multiplied by a 
factor of $a/a'$. The same for $l(k)$ and $\tau(k)$, but in addition, $l(k)$ and $\tau(k)$ 
will now be divided by the same factor since these two quantities involve the derivative 
with respect to $k$. The data obtained for Well~III in Table~\ref{Table1} confirms this 
scaling. Note that the centers of the peaks and the poles themselves are scaled by 5 and 
the peak of the effective traversal length $l(k^*)$ divided by $2a$ is the same for both 
wells. Here, $k^*$ are the values for which $l(k)$ is a local maxima.
   
Often one finds in the literature approximations or statements made for deep potentials. 
The abovementioned scaling emphasizes that the more relevant quantity is $\alpha$
(\ref{eq.3p1b}), and not just the depth. A peculiarity of the square well is that the 
value of $\alpha$ determines the number of bound states and hence, very strongly, the 
scattering properties for the low lying resonances. We have found that a very good 
approximation for the number $N_B$ of bound states is given by the integer part of
\cite{formula}
\begin{equation}
\label{eq.3p17}
Q_B= (\alpha/\pi +1/2) \ ,
\end{equation}
In column 4 of Table~\ref{Table1}, we list $Q_B$ for all  wells considered here. The 
formula predicts that there are 3 bound states for Wells~I and III, 17 bound states 
for Well~II, 12 for Well~IV, and so on. Moreover, the fractional part of $Q_B$ indicates us 
how close the bound state is to the continuum. In other words, it indicates how close it is 
to fit the next bound state. Let us skip momentarily Well~IV and consider Well~V listed in 
Table~\ref{Table1}; Well~V has $\alpha=39.2505$ and its scattering functions are plotted 
in Fig.~\ref{fig3}. According to our interpretation above, Well~V (with $Q_B=12.994$) is 
almost strong enough to bind 13 states. I.e., the 13th state corresponds to the first 
scattering resonance close to $k=0$. The predictions based on the interpretation of  
formula (\ref{eq.3p17})  have been verified numerically. We shall now see how critically 
the fractional part of $Q_B$ determines the scattering properties of the well. 

Notice that the real part of the first resonance pole for Well~V (II) is the smallest (largest) 
of the first five wells and how this reflects in the scattering properties displayed in 
Figs.~\ref{fig1} to \ref{fig3}. Focusing on Well~V, Fig.~\ref{fig3}, we see that now the 
first resonance of all its scattering functions come closer to $k=0$ as predicted above. We 
also note that the resonance profiles have changed drastically. The bell-like shape of 
$\sigma_{\phi}(k)$ and of $P(k)$ have become skewed and become zero at $k=0$. Now 
the maxima of $l(k)$ is significantly larger than $2a$ but  has lost its bell shape profle. 
Moreover $\tau(k)\rightarrow \infty$ as $k\rightarrow 0$.

We can understand this behavior by examining the formulas (\ref{eq.3p12}) to 
(\ref{eq.3p14b}).  These show that $|A|^2\rightarrow 0$ and $P(k)$ clearly zero at 
$k=0$. Further $l(k)/(2a)$ equals $(qa/\alpha^2)\tan(qa)$. 
Note also that now the local maxima of $\sigma_{\phi}(k)$, $P(k)$, and even $l(k)$, are 
more distant now from the real part of the $S$-matrix pole. This feature, discussed in a 
future publication, can be understood  by means of reaction matrix theory.
  
The resonance patttern displayed by Well~V pertains to very low energy resonances, not 
the case of the so-called zero energy resonances where the $s$-wave scattering cross 
section (\ref{sigmak2}) becomes infinite, see 
e.g.~\cite{Taylorbook,Newtonbook,ABohmbook,Meyer76}.
   
Now we want to show numerically that all resonances of $l(k)$ and $\tau(k)$ yield a 
positive time delay. Perhaps ``positive time delay" seems redundant for if there is a time 
delay, it must be positive. However, since time delay is already the name of the function 
and it can be zero or negative, we add the adjective positive. Figure~\ref{fig4} shows the 
{\it first} maxima of $l(k)/(2a)$ as a function of $\alpha$.  Recall that $k^*$ are the 
values for which $l(k)$ is a local maxima. Inspection of this plot  and its data shows that all 
first maxima $l(k^*)$ are larger or equal to $2a$, hence $\tau(k^*)$ is greater than zero. 
Notice the periodic monotonic increase of $l(k*)/(2a)$ as $\alpha$ increases and then 
drops back to $l(k*)/(2a)\approx 1$. In the inset of Fig.~\ref{fig4} we zoom on four
consecutive first maxima of $l(k)$, corresponding to 4 wells varying slightly in their values 
of $\alpha$. These four wells are labeled IV, V, VI, and VII in Table~\ref{Table1}. Wells~IV, 
V, and VI, with $\alpha=39.0535$, 39.1520, and 39.2505, respectively, have 12 bound 
states according to the integer value of their $Q_B$. By the same token, Well~VII has 13 
bound states. The effective traversal distance and the time delay of these four wells are 
plotted in Fig.~\ref{fig5}. Figure~\ref{fig5} exemplifies the pattern observed in 
Fig.~\ref{fig4}; namely, as $\alpha$ increases an additional state is bound; the closer the 
first resonance state is to $k=0$, the larger the time delay. Simultaneously, the value of 
the resonant part of the phase shift $\phi$ moves down away from $\pi/2$ (see 
Table~\ref{Table1}). Conversely, as the first maxima of $l(k)$ moves up away from $k=0$ 
the phase $\phi$ at the local maxima of $l(k)$ becomes closer to $\pi/2$. The larger 
$\alpha$ is, the sharper the resonances are and  at the same time the closer these occur 
to $\pi/2$. We have shown that $\tau(k)=0$ exactly at $\phi=\pi/2$. This is perhaps the 
reason why in previous studies, which assume very large values of $\alpha$ no time delays 
have been observed.

\section{Conclusions}

The reason for requiring, in the established definition of a scattering resonance, a rapid 
increase in the total phase shift $\theta$ comes from the idea that a ``true" resonance is 
to correspond to the formation of a meta-stable state or to the temporary trapping of the 
projectile in the scattering region (see e.g.~\cite{Taylorbook,Newtonbook,Joachainbook}). 
Trapping leads to a time delay of the scattered wave, which in turn implies that the 
Wigner-Smith time delay $\tau(k)$ must be positive. 

Our expressions shows that for $\tau(k)$ to be greater than zero, the effective traversal 
distance $l(k)$ must be greater than twice the width of the well. We demonstrated 
numerically that this is so for the square well for all the local maxima of $\tau(k)$.
Moreover, we showed numerically that the peaks of $\tau(k)$ and $l(k)$ occur very close 
to each other and to the real part of the $S$-matrix resonance poles. We also showed that 
although the peaks of $\tau(k)$ occur for values of $\phi$ slightly smaller than $\pi/2$ 
(modulo $\pi$),  $\phi$ jumps by about $\pi$  from begining to end of the resonance 
passing through $\pi/2$. So, it meets the requirements of the stablished definition or 
resonance, except that the peak of $\tau(k)$ does not occur exactly at $\pi/2$, where 
the resonant part of the scattering cross section reaches its unitary limit. On the other 
hand, we showed that as $\alpha$ becomes large, the local maxima of $\tau(k)$ and 
$l(k)$ occur closer to $\phi=\pi/2$ (modulo $\pi$). Simultaneously  $l(k)\rightarrow 2a$ 
and consequently $\tau(k) \rightarrow 0$. This explains why most studies on the 
resonances properties of the square well, which focus on large values of the ``strength'' 
$\alpha$ (usually 
referrred to as deep potentials), arrive at the conclusion that the square well does not 
produce meta-stable states.  

The study of this simple system leads us to question about the conventional or popular 
definition of scattering resonances. Specifically, it raises the question about how essential 
is it that the time delay be positive exactly at $\pi/2$ for it to be true scattering 
resonance? Why is the cross section and not other scattering function, like the dwelling 
time, or the trapping probability be the standard or preferred scattering function 
to define a scattering resonance? After all, time delay $\tau(k)$, trapping probability 
$P(k)$, relative intensity $|A|^2$, and cross secction $\sigma(k)$ are all aspects of 
the same scattering phenomena and these are expected to display similar behavior under 
resonance. $P(k)$ and $\sigma(k)$ have more in common with each other than with 
$\tau(k)$ and $l(k)$. The first pair relate to probabilities while the last pair relate to 
changes. So, it appears that the information they carry is complementary.

The quantum mechanical definition ultimately originates from the classical definition of 
scattering cross section. Is this the main reason why even popular text books in quantum 
mechanics at graduate level define scattering purely in terms of the cross section?

We believe these questions have not been raised, or at least, as far as we know, have not 
been published, because most studies of resonances deal with sufficiently sharp 
resonances, for which, as we have seen, time delay, and cross section coincide in their 
peaks. In fact, it is well known that both, the cross section and the time delay, are 
proportional to each other and are described by the Breit-Wigner resonance formula. This 
of course is an approximation, valid for sharp resonances, or equivalently for resonance 
poles near the real axis where a one-pole aproximation lead to the Breit-Wigner formula. 
The center of the resonance is usually associated with the real part of the pole in energy 
space. Our numerical studies show, however, that it is more closely related to the absolute 
value of the resonance pole (in energy or wave number space).

It is natural to ask if the scattering properties presented here for the central square well
are expected to be true also for more general or realistic finite-range potentials. The main
results rely on the separation of the phase shift into the hard-sphere phase shift $-ka$ and
resonance part $\phi$. If the radius of interaction $a$ is well known, then the choice 
$r=a$ is the optimum choice. If $a$ is only known approximately or the evaluation 
(measurement) of the phase shift can be done only for a radius $r>a$ then it is necessary
to examine how the scattering quantities $\tau(k)$, $l(k)$, $P(k)$ and $\sigma_\phi(k)$
change with changes of $r$ (in the asymptotic region). First we note that the full phase
shift $\theta$ does not depend on $r\geq a$. Thus, the time delay, which is ultimately
defined in terms of $\theta$, does not depend on $r\geq a$ either. The same is true 
for the scattering cross section $\sigma_\theta(k)$.

For the speed of the phase shift $l(k)$ let us consider two arbitrary radius of separation
($r_1$ and $r_2$) for the resonant part $\phi$; $\phi_1=\theta+kr_1$ and 
$\phi_2= \theta+kr_2$. Since $\theta$ is independent of the radius $r$, it follows that
$\phi_2=\phi_1+k(r_2-r_1)$; hence $l_2(k)=l_1(k)+(r_2-r_1)$. Thus these two differ 
by the constant $r_2-r_1$. Hence the shape of $l(k)$ evaluated at some $r\geq a$ is 
the same (except higher by a distance $r-a$) as if evaluated at the radius of interaction 
$a$. Moreover, the centers of their local maxima occur at the same value of $k$. This
result is independent of the potential, as long as it has a finite range. Further, for the
central square well we showed numerically that the $n^{th}$ local maxima of $l(k)$ 
occur in a very small neighborhood of $\kappa_n$, the real part of the pole in complex 
$k$-space. Work in progress suggests strongly that this feature is true for any finite-range
potential.

This property is not shared by the trapping probability. Consider the integrals 
$P_a= \int_0^{a} |\psi(k;r)|^2 dr$ and $P_r=\int_0^{r} |\psi(k;r)|^2dr$,
corresponding, respectively, to radii $a$, the actual radius of interaction, and $r>a$. 
It can be easily shown that 
$P_r=(a/r)P_a + 2(r-a)/r-[\sin(2\phi_r)-\sin(2\phi_a)]/kr$, where 
$\phi_r$ and $\phi_a$ are the resonant part of the phase shift evaluated at radius 
$r>a$ and $a$, respectively. This shows that the profiles of $P_a$ and $P_r$ as a
function of $k$ can be very different if $r-a$ is not very small or $k$ is not very large. 
In such case, the centers of their peaks will be very different too. On the other hand, 
Figs.~1-3, for the central square well, show that when $r=a$ the trapping probability
$P_a(k)$ under resonance conditions assumes a lorentzian-like shape, similar to the 
shape of the other scattering quantities. If this lorentzian shape were to be universal 
for all short-range potentials, then it can be used as a criteria to determine the radius 
of interactions. Work in progress carried out in a variety of short-range potentials indicates
that indeed this shape is universal. 

This universality is not obeyed for the purely resonant scattering cross section 
$\sigma_\phi(k)$. In fact, even for the central square well, different values of $r$ used in 
the evaluation of the resonant phase $\phi$ can produce shapes of $\sigma_\phi(k)$
that differ radically from the well known Breit-Wigner resonance formula, attained when
$r=a$. However, this may not be seen as a negative quality for it serves to discriminate
between the actual radius of interaction and other radii and is consistent with the behavior
of the trapping probability. 

Thus, while not all features presented by the central square well can be generalized to
arbitrary potentials, the merit of the square well is that it reveals subtle but important
distinctions between the various scattering functions that motivates their study in more
realistic systems.\\

{\bf Acknowledgments}.
This work was partially supported by Fondo Institucional PIFCA 
(Grant No.~BUAP-CA-169). 
J.A.M.-B thanks support from CONACyT (Grant No.~CB-2013/220624)
and VIEP-BUAP (Grant No.~MEBJ-EXC14-I).
A.A.F.-M. thanks financial support from DGAPA-UNAM.


\begin{thebibliography}{00}

\bibitem{Workman2013}
R. L. Workman, L. Tiator, and A. Sarantsev,
Phys. Rev. C {\bf 87}, 068201 (2013).

\bibitem{Descouvemont2010} 
P. Descouvemont and D. Baye, 
Rep. Prog. Phys. {\bf 73}, 036301 (2010).

\bibitem{Ceci2013} 
S. Ceci, M. Korolija, and B. Zauner, 
Phys. Rev. Lett {\bf 111}, 112004 (2013).

\bibitem{BDietz2010}
See, e.g., 
B. Dietz, H. L. Harney, A. Richter, F. Sch\"afer, and H. A. Weidenm\"uller, 
Phys. Lett. B {\bf 685}, 263 (2010); 
G. Shchedrin and V. Zelevinsky, 
Phys. Rev. C. {\bf  86}, 044602 (2012); and 
C. Poli, G. A. Luna-Acosta, and H.-J. St\"ockmann, 
Phys. Rev. Lett. {\bf 108}, 174101 (2012).

\bibitem{Carvalho2002}
C. A. A. de Carvalho and H. M. Nussenzveig, 
Phys. Rep. {\bf 364}, 83 (2002).

\bibitem {Taylorbook}
J. R. Taylor, 
{\em Scattering Theory} (John Wiley and Sons, New York, 1972).

\bibitem{Newtonbook} 
R. G. Newton, 
{\em Scattering Theory of Waves and Particles}, 2nd ed. 
(Springer-Verlag, New York, 1982).

\bibitem{ABohmbook}
A. B\"ohm, 
{\em Quantum Mechanics. Foundations and Applications} 
(Springer Science + Business Media, New York, 1986).

\bibitem{PDG2014}
K. A. Olive, et. al. (Particle Data Group), 
Chin. Phys. C {\bf 38}, 090001 (2014).

\bibitem{LL65}
L. D. Landau and E. M. Lifshitz,
{\em Quantum Mechanics: Non-Relativistic Theory}, Vol. 3 (2nd ed.)
(Pergamon Press, 1965).

\bibitem{Simon00}
B. Simon,  
J. Funct. Anal. {\bf 178}, 396 (2000).

\bibitem{Weberetal1982} 
T. A. Weber, C. L. Hammer, and V. S. Zidell, 
Am. J. Phys. {\bf 50}, 839 (1982).

\bibitem{Nussenzvieg59} 
H. M. Nussenzvieg, 
Nucl. Phys. {\bf 11}, 499 (1959).

\bibitem{Lane86}
A. M. Lane, 
J. Phys. B: At. Mol. Phys. {\bf 19}, 253 (1986).

\bibitem{Vogt1962}
E. W. Vogt, 
Rev. Mod. Phys. {\bf 34}, 723 (1962);
Phys. Lett. B {\bf 389}, 637 (1996). 

\bibitem{BlattWeisskopf} 
J. M. Blatt and V. F. Weisskopf, 
{\em Theoretical Nuclear Physics} 
(Dover Publications, New York, 1991).

\bibitem{Meyer76}
H.-D. Meyer and K. T. Tang, 
Z. Physik {\bf 279}, 349 (1976).

\bibitem{BurkeAMOBook2011}
P. G. Burke,
{\em R-Matrix Theory of Atomic Collisions: Application to Atomic, Molecular and Optical Processes} 
(Springer, Berlin, 2011).

\bibitem{Vanroose97} 
See, e.g.,
J. Okolowicz, M. Ploszajczak, and I. Rotter,
Phys. Rep. {\bf 374}, 271 (2003) 
and
W. Vanroose, P. Van Leuven, F. Arickx, and J. Broeckhove, 
J. Phys. A: Math. Gen. {\bf 30}, 5543 (1997).

\bibitem{Fano61}
U. Fano, 
Phys. Rev. {\bf 124}, 1866 (1961).

\bibitem{Miroshnichenko10} 
A. E. Miroshnichenko, S. Flach, and Y. S. Kivshar, 
Rev. Mod. Phys. {\bf 82}, 2257 (2010).

\bibitem {Joachainbook}
C. J. Joachain, 
{\em Quantum Collision Theory} 
(North-Holland, Amsterdam,1987).

\bibitem{Wigner1955}
E. P. Wigner, 
Phys. Rev. {\bf 98}, 145 (1955).

\bibitem{TSmith60}
F. T. Smith,
Phys. Rev. {\bf 118}, 349 (1960).

\bibitem{Amrein07}
W. O. Amrein and P. Jacquet,
Phys. Rev. A {\bf75}, 022106 (2007).

\bibitem{EPE47}
E. P. Wigner and L. Eisenbud,
Phys. Rev. {\bf 72}, 29 (1947).

\bibitem{LaneThomas58}
A. M. Lane and R. G. Thomas, 
Rev. Mod. Phys. {\bf 30}, 257 (1958).

\bibitem{PMbook} 
P. A. Mello and N. Kumar,
{\em Quantum Transport in Mesoscopic Systems}
(Oxford University Press, New York, 2004). 

\bibitem{Klaiman2010}
S. Klaiman and N. Moiseyev,
J. Phys. B: At. Mol. Opt. Phys. {\bf 43}, 185205 (2010).

\bibitem{Anisovich2012} 
A. V. Anisovich, R. Beck, E. Klempt, V. A. Nikonov, A. V. Sarantsev, and U. Thoma, 
Eur. Phys. J. A {\bf 48}, 15 (2012).

\bibitem{formula} 
This expresion can be obtained really easily by assuming that the bound states and the 
scattering states of the well obey Neumann boundary conditions at $r=a$. Then, 
$qa=(2n-1)\pi/2$. Using (\ref{eq.3p1a}) with $k=0$ and solving for $n\equiv N_B$
yields (\ref{eq.3p17}).

\end{thebibliography}
\end{document}